%Paper: hep-th/9508099
%From: G. Mussardo <mussardo@mersenne.sissa.it>
%Date: Mon, 21 Aug 95 13:20:12 METDST
%Date (revised): Thu, 24 Aug 95 13:10:21 METDST

%%%%%%%%%%%%%%%%%%%%%%%%%%%%%%%%%%%%%%%%%%%%%%%%%%%%%%%%%%%%%%%%%%%%%
%%%%%%%%%%%%%%%%%%%%%%%%%%%%%%%%%%%%%%%%%%%%%%%%%%%%%%%%%%%%%%%%%%%%%
%%%%%%%%%%%%%%%%%%%%%%%%%%%%%%%%%%%%%%%%%%%%%%%%%%%%%%%%%%%%%%%%%%%%%
\input harvmac.tex      %% input your version
\input epsf.tex         %%        "
%\input harvmac
%%%%%%%%%%%%%%%%%%%%%%%%%%%%%%%%%%%%%%%%%%%%%%%%%%%%%%%%%%%%%%%%%%%%%
%%%%
%%%%
%%%%        On Ising Correlation Functions with Boundary
%%%%            Magnetic Field
%%%%
%%%%
%%%%               By R. Konik, A. LeCLair, and G. Mussardo
%%%%
%%%%
%%%%%%%%%%%%%%%%%%%%%%%%%%%%%%%%%%%%%%%%%%%%%%%%%%%%%%%%%%%%%%%%%%%%%
%%%%%%%
%%%%            Macros needed:  harvmac.tex   (from bulletin board)
%%%%
%%%%
%%%%%%%%%%%%%%%%%%%%%%%%%%%%%%%%%%%%%%%%%%%%%%%%%%%%%%%%%%%%%%%%%%%%%
%%%%%%%%%

%%%%%%%%%%%%%%%%%%%%%%%%%%%%%%%%%%%%%%%%%%%%%%%%%%%%%%%%%%%%%%%
%
%		DEFINITIONS FOR TEX
%
%%%%%%%%%%%%%%%%%%%%%%%%%%%%%%%%%%%%%%%%%%%%%%%%%%%%%%%%%%%%%%%
%

\def\tilde{\widetilde}
\def\bar{\overline}
\def\kap{\kappa}
\def\hat{\widehat}
\def\*{\star}
\def\[{\left[}
\def\]{\right]}
\def\({\left(}		
\def\){\right)}

%
%%%%%%%%%%%%%%%%%%%%%%%%%%%%%%%%%%%%%%%%%%%%%%%%%%%%%%%%%%%%%%%
%
\def\zb{{\bar{z} }}
\def\frac#1#2{{#1 \over #2}}
\def\inv#1{{1 \over #1}}

\def\d{\partial}

\def\2pi{\hbox{$2\pi i$}}

\def\dsl{\raise.15ex\hbox{/}\kern-.57em\partial}
\def\Dsl{\,\raise.15ex\hbox{/}\mkern-.13.5mu D}
%
%%%%%%%%%%%%%%%%%%%%GREEK LETTERS%%%%%%%%%%%%%%%%%%%%%%%%%%%%%%
%

\def\be{\beta}
\def\al{\alpha}

\def\sig{\sigma}

%
%%%%%%%%%%%%%%%%%%%CALIGRAPHIC LETTERS%%%%%%%%%%%%%%%%%%%%%%%%%
%

%

\def\2pi{\hbox{$2\pi i$}}

\def\dsl{\raise.15ex\hbox{/}\kern-.57em\partial}
\def\Dsl{\,\raise.15ex\hbox{/}\mkern-.13.5mu D}
%
%%%%%%%%%%%%%%%%%%%%GREEK LETTERS%%%%%%%%%%%%%%%%%%%%%%%%%%%%%%
%
%%%%%%%%%%%%%%% MATH CHARACTERS %%%%%%%%%%%%%%%%%%%%%%%%%%%%
%
\font\numbers=cmss12
%\font\numbers=cmu10 scaled\magstep1
\font\upright=cmu10 scaled\magstep1
\def\stroke{\vrule height8pt width0.4pt depth-0.1pt}
\def\topfleck{\vrule height8pt width0.5pt depth-5.9pt}
\def\botfleck{\vrule height2pt width0.5pt depth0.1pt}
\def\Zmath{\vcenter{\hbox{\numbers\rlap{\rlap{Z}\kern
0.8pt\topfleck}\kern
2.2pt
                   \rlap Z\kern 6pt\botfleck\kern 1pt}}}
\def\Qmath{\vcenter{\hbox{\upright\rlap{\rlap{Q}\kern
                   3.8pt\stroke}\phantom{Q}}}}
\def\Nmath{\vcenter{\hbox{\upright\rlap{I}\kern 1.7pt N}}}
\def\Cmath{\vcenter{\hbox{\upright\rlap{\rlap{C}\kern
                   3.8pt\stroke}\phantom{C}}}}
\def\Rmath{\vcenter{\hbox{\upright\rlap{I}\kern 1.7pt R}}}
\def\Z{\ifmmode\Zmath\else$\Zmath$\fi}
\def\Q{\ifmmode\Qmath\else$\Qmath$\fi}
\def\N{\ifmmode\Nmath\else$\Nmath$\fi}
\def\C{\ifmmode\Cmath\else$\Cmath$\fi}
\def\R{\ifmmode\Rmath\else$\Rmath$\fi}
%%%%%%%%%%%%%%%%%%%%%%%%%%%%%%%%%%%%%%%%%%%%%%%%%%%%%%%%%%%%%%%%%
 %%%%%%%%%%%%%%%%%% END OF DEFINITIONS %%%%%%%%%%%%%%%%%%%%%%
 %%%%%%%%%%%%%%%%%%%%%%%%%%%%%%%%%%%%%%%%%%%%%%%%%

\Title{\vbox{\baselineskip12pt
\hbox{CLNS 95/1355}
\hbox{ISAS/EP/95/93}}}
{\vbox{\centerline{On Ising Correlation Functions}
\centerline{with Boundary Magnetic Field} }}

\bigskip
\bigskip

\centerline{Robert Konik and Andr\'e  LeClair}
\medskip\centerline{Newman Laboratory}
\centerline{Cornell University}
\centerline{Ithaca, NY  14853}
\bigskip\bigskip
\centerline{and}
\bigskip
\centerline{Giuseppe Mussardo}
\medskip\centerline{International School for Advanced Studies and INFN}
\centerline{34014 Trieste}

\vskip .3in

{\bf Abstract}: Exact expressions of the boundary state and the form factors
of the Ising model are used to derive differential equations for the
one-point functions of the energy and magnetization operators of the model
in the presence of a boundary magnetic field. We also obtain explicit
formulas for the massless limit of the one-point and two-point functions of the
energy operator.

\Date{8/95}
%\draftmode
%
%
%
%
%
%
%      pecdef

\noblackbox

\def\zb{{\bar{z}}}

\def\eh{\hat{e}}
\def\fh{\hat{f}}
\def\dh{\hat{D}}
\def\ct{\tilde{C}}
\def\dt{\tilde{D}}

%
%
%
%
%sample reference
%
%on an operator formulation of the superstring\ref\a{\sdual} .
%
%sample equations
%\eqn\one{
%\V 123 A\rangle_1 \vb_2 \vc_3 =
%\langle h_1 \left[ V_A (0)\right] h_2 \left[ V_B (0) \right]
%h_3 \left[ V_C (0) \right] \rangle . }
%
%
% \eqn\two{\eqalign{  a & = b \cr c & = d \cr}}
%
%\eqnn\two
%$$\eqalignno{
% h_1 (z ) &= z \cr
 %h_2 (z ) &= { 1\over {1-z } }&\two\cr
% h_3 (z ) &= { {z -1}\over {z} } .\cr}$$
%
%eq a,b,c etc
%\eqna\three
%$$\eqalignno{
%.............&\three {a} \cr
%..........&\three {b} \cr
%}$$
%\eqn\number{\eqalign{  ......... \cr}}       numbers automatically
%\newsec{Introduction }
%\centerline{Acknowlegements}
%\figures
%\fig{1}{bla}
%\appendix{A}{blaaaaaaaaaaaaa}
%\listrefs
%\end

\newsec{Introduction}

Quantum Field Theories with boundary conditions
\ref\binder{K. Binder, in
{\it Phase transitions and critical phenomena}, vol. 8, ed. C. Domb and
J. Lebowitz (Academic Press, London 1983).}
\ref\cardy{J.L. Cardy,
Nucl. Phys. B 240 [FS12] (1984), 514; Nucl. Phys. B 324 (1989), 581.}
\ref\cher{I. Cherednik, Theor. Math. Phys. 61 (1984), 997.}
\ref\gz{S. Ghoshal and A. Zamolodchikov,
Int. Journ. Mod. Phys. A 9 (1994), 3841.}
\ref\kf{A. Fring and R. Koberle, Nucl. Phys. B 419 (1994), 647; Nucl. Phys.
B 421 (1994), 159.}
have interesting applications to quite a large class of physical phenomena,
among them the Kondo problem
\ref\al{I. Affleck and A. Ludwig, Nucl. Phys. B 360 (1991), 641;
Phys. Rev. B 48 (1993), 7297.}, absorption of polymers
on a surface \ref\pol{E. Eisenriegler, {\it Polymers Near Surfaces}
(World Scientific Publishing, 1993).} \ref\polymer{P. Fendley and H. Saleur,
J. Phys. A (1994), L789.} and transport properties of Luttinger liquids
\ref\lut{P. Fendley, A. Ludwig and H. Saleur,
{\it Exact non-equilibrium transport through point contacts in
quantum wires and fractional quantum Hall effect}, preprint UCS-95-007.}.
In addition to extensive quantities which can be
computed by means of the Bethe Ansatz \ref\chattba{R. Chatterjeee,
Mod. Phys. Lett. A 10 (1995), 937.} \ref\btba{A. LeClair, G. Mussardo,
H. Saleur and S. Skorik, {\it Boundary energy and boundary states in
integrable quantum field theories}, hep-th/9503227, to appear on
Nucl. Phys. B.}, it is also interesting to analyse the behaviour of local
observables in the presence of boundary conditions. In this paper we discuss
a simple example of QFT with boundary conditions which has the advantage of
being suitable for an exact analysis. The example we consider is the
computation of one-point functions of the local operators of the Ising model
defined in an half-space. Original lattice derivation of these quantities may
be found in the refs. \ref\mccoy{B.M. McCoy and T.T. Wu, Phys. Rev. 162
(1967) 436;
Phys. Rev. 174 (1968) 546.}\ref\mccoyb{B.M. McCoy and T.T. Wu, The Two
Dimensional Ising Model, Harvard University Press, Cambridge 1973.}
\ref\bariev{R.Z. Bariev, Teor. Mat. Fiz. 40 (1979),
40; Teor. Mat. Fiz. 42 (1980), 262; Teor. Mat. Fiz. 77 (1988), 1090.}.
Before proceeding in our computation, let us briefly recall some general
aspects of the problem.

The best way to compute correlation functions of a QFT with boundary is
to take full advantage of the solution of the theory in the bulk and of the
knowledge of the boundary state \gz
\ref\defect{G. Delfino, G. Mussardo and P. Simonetti,
Nucl. Phys. B 432 [FS] (1994), 518.}. Namely, instead of considering
the boundary placed at the spatial coordinate $x=0$, it is most convenient
to consider the boundary placed at $t=0$ and described by the boundary state
$\mid B>$ (Figure 1).

%%%%%%%%%%%%%%%%%%%  fig 1  %%%%%%%%%%%%%%%%%%%%%
\midinsert
\epsfxsize = 3in
\bigskip\bigskip\bigskip\bigskip
\vbox{\vskip -.1in\hbox{\centerline{\epsffile{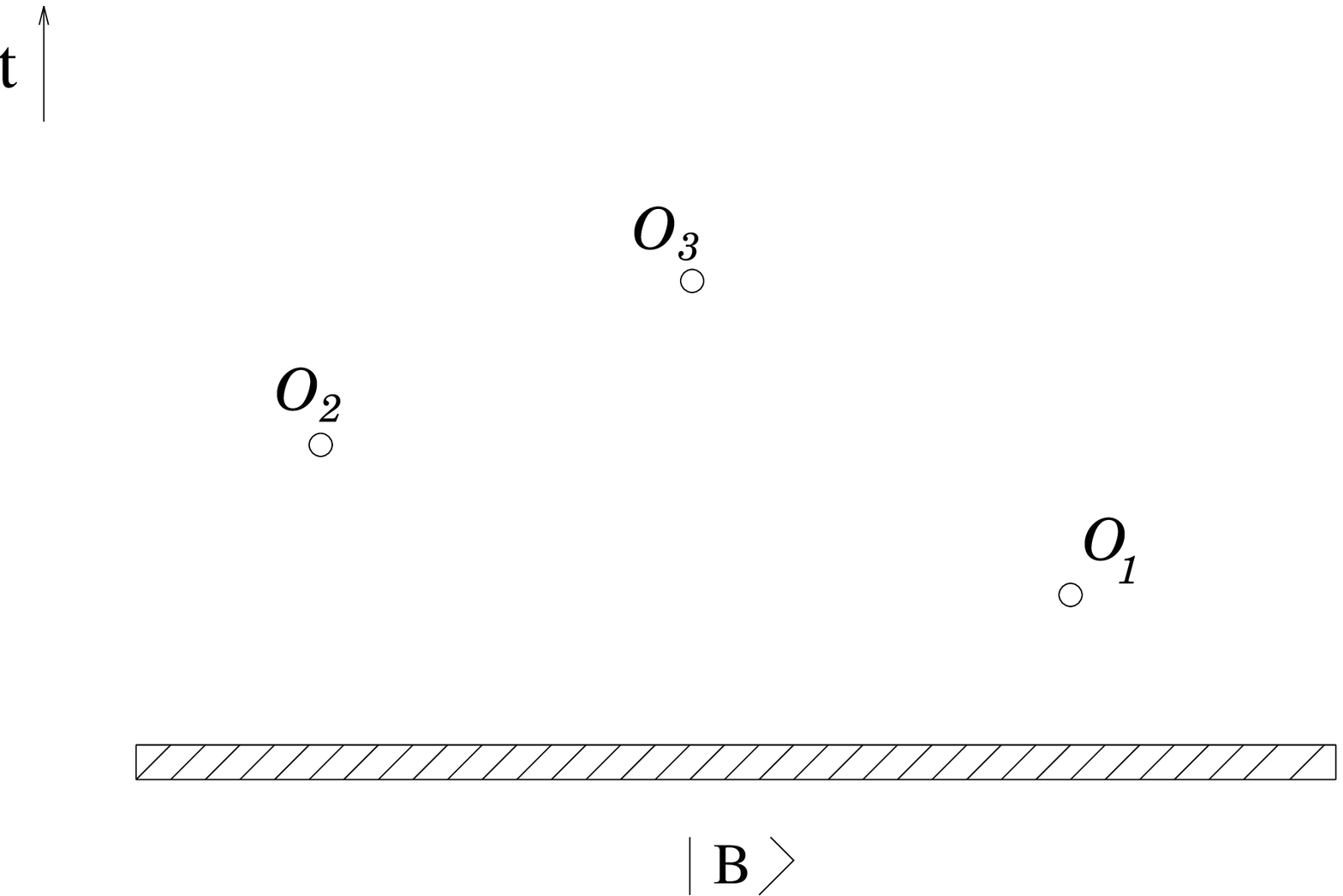}}}
\vskip .3in
{\leftskip .5in \rightskip .5in \noindent \ninerm \baselineskip=10pt
\leftskip 1in Figure 1. Boundary geometry for the computation of correlators.
\smallskip}} \bigskip
\endinsert
%%%%%%%%%%%%%%%%%%% End fig 1   %%%%%%%%%%%%%%%%%%%%%%%

Then, the correlation functions can be expressed as
\eqn\eIi{
<{\cal O}_1(x_1,t_1) \ldots {\cal O}_n(x_n,t_n)> \,=\,
\frac{
<0\mid T_t\left[{\cal O}_1(x_1,t_1) \ldots {\cal O}_n(x_n,t_n)\right]\mid B>}
{<0 \mid B>}
. }
In this geometry, the Hilbert space of the theory is as in the bulk and
therefore, even in the presence of the boundary line, the local operators
${\cal O}_i$ can be completely characterized by the bulk Form Factors
$<\beta_1,\ldots,\beta_n\mid {\cal O}_i\mid \beta_{n+1},\ldots,\beta_m>$,
where $\beta_i$ are the rapidity variables, related to the two-dimensional
momenta by $p_i^{(0)} = m_i \cosh\beta_i$, $p_i^{(1)} = m_i \sinh\beta_i$.

Let us focalize our analysis to the Ising model with a boundary in the low
temperature phase ($T < T_c$) \gz. The system can be described in terms of
massive fermionic operators $A(\beta)$ and $A^{\dagger}(\beta)$ with
the usual anti-commutation relations. The mass $m$ is a linear measurement
of the deviation of the temperature with respect to the critical one,
$m = 2\pi (T_c - T)$. The spins on the boundary can be subjected to three
possible boundary conditions, namely (a) they can be frozen to one of the
fixed values $\pm 1$  (``fixed boundary condition"), (b) they can be
completely free to fluctuate (``free boundary condition") or (c)
they can be coupled to a boundary magnetic field $h$ (``magnetic boundary
condition")\foot{In order to simplify the notation, from now on
we denote the free b.c. with the index $(+)$, the fixed b.c. with the index
$(-)$ and the magnetic b.c. with $(h)$.}. The detailed analysis of these
three possibilities is given in the original reference \gz,
and here we simply
recall the results for the reflection $S$-matrix $R(\beta)$ associated to each
of them. For the free boundary condition we have
$
R_{+}(\beta) \,=\,i \coth \left(\frac{\beta}{2} -\frac{i \pi}{4}\right) \,
$,
for the fixed boundary condition
$
R_{-}(\beta) \,=\,-i \tanh \left(\frac{\beta}{2} -\frac{i \pi}{4}\right)
$,
and finally for the magnetic boundary condition
\eqn\emag{
R_{h}(\beta) \,=\,-i \tanh \left(\frac{\beta}{2} -\frac{i \pi}{4}\right)\,
{{\kap - i \sinh\beta}\over{\kap + i \sinh\beta}}
\,\, ,} where
$
\kappa\,=\, 1 - \frac{h^2}{2m}
$.
Varying $h$, we can interpolate between the free boundary condition
($h = 0$) and the fixed one ($h \rightarrow \infty$).

Given an amplitude $R(\beta)$, the corresponding boundary state can be then
expressed as\foot{
The most general expression of the boundary state may
also involve an additional term relative to the zero mode. Here we discard it
since it does not enter our following computations.}
\eqn\eIvi{
\mid B> \,=\, \exp\left[\int_0^{\infty} \frac{d\beta}{2\pi}
\hat R(\beta) A^{\dagger}(-\beta) A^{\dagger}(\beta)\right] \mid 0>
\,\, ,}
where $\hat R(\beta) = R\left(\frac{i \pi}{2} -\beta\right)$.
The structure of the boundary state $|B \rangle$ is such that
the states entering its definition consist of pairs of particles
with opposite rapidity (Cooper pairs).

We can now use the Form Factors of the Ising model determined in
\ref\kw{B. Berg, M. Karowski and P. Weisz, Phys. Rev. D 19 (1979), 2477.}
\ref\zcm{V.P. Yurov and Al.B. Zamolodchikov, Int. J. Mod. Phys. A 6
(1991), 3419; J.L. Cardy and G. Mussardo, Nucl. Phys. B 340 (1990), 387.}
and the expression of the boundary state \eIvi\
to compute the correlation functions in the presence of the boundary.

\vskip .3in

\newsec{One-point and two-point functions of the energy operator}

The simplest correlation function is the one-point function of the
energy operator $\epsilon(x,t)$ which can be computed through the formula
\eqn\eIvii{
\epsilon_0(t)\,=\,\sum_{n=0}^{\infty} <0\mid\epsilon(x,t)\mid n>
<n\mid B>\,\,\,.
}
The energy operator couples the vacuum only to the two particle state
and for its matrix element in the euclidean space we have
\eqn\eIviii{\eqalign{
<0\mid \epsilon(x,t)\mid \beta_1,\beta_2>\,&=\, - 2 \pi m \, i\,\,
\sinh\frac{\beta_1-\beta_2}{2}  \cr
&~~~~~~~
\times \,\exp\left[-m t\, (\cosh\beta_1 + \cosh\beta_2) +
i m x \,(\sinh\beta_1 + \sinh\beta_2)\right]\,\, .
\cr}}
Hence the sum \eIvii\ consists of only one term and the one-point function
of the energy operator can be expressed
as
\eqn\eIix{
\epsilon_0(t)\,=\ -i \, m \,\int_0^{\infty} d\beta
\sinh\beta \,\hat R(\beta) \,e^{-2mt\cosh\beta} \,\,\, .
}
Its graphical representation is given in Figure 2.

%%%%%%%%%%%%%%%%%%%  fig 2  %%%%%%%%%%%%%%%%%%%%%
\midinsert
\epsfxsize = 3in
\bigskip\bigskip\bigskip\bigskip
\vbox{\vskip -.1in\hbox{\centerline{\epsffile{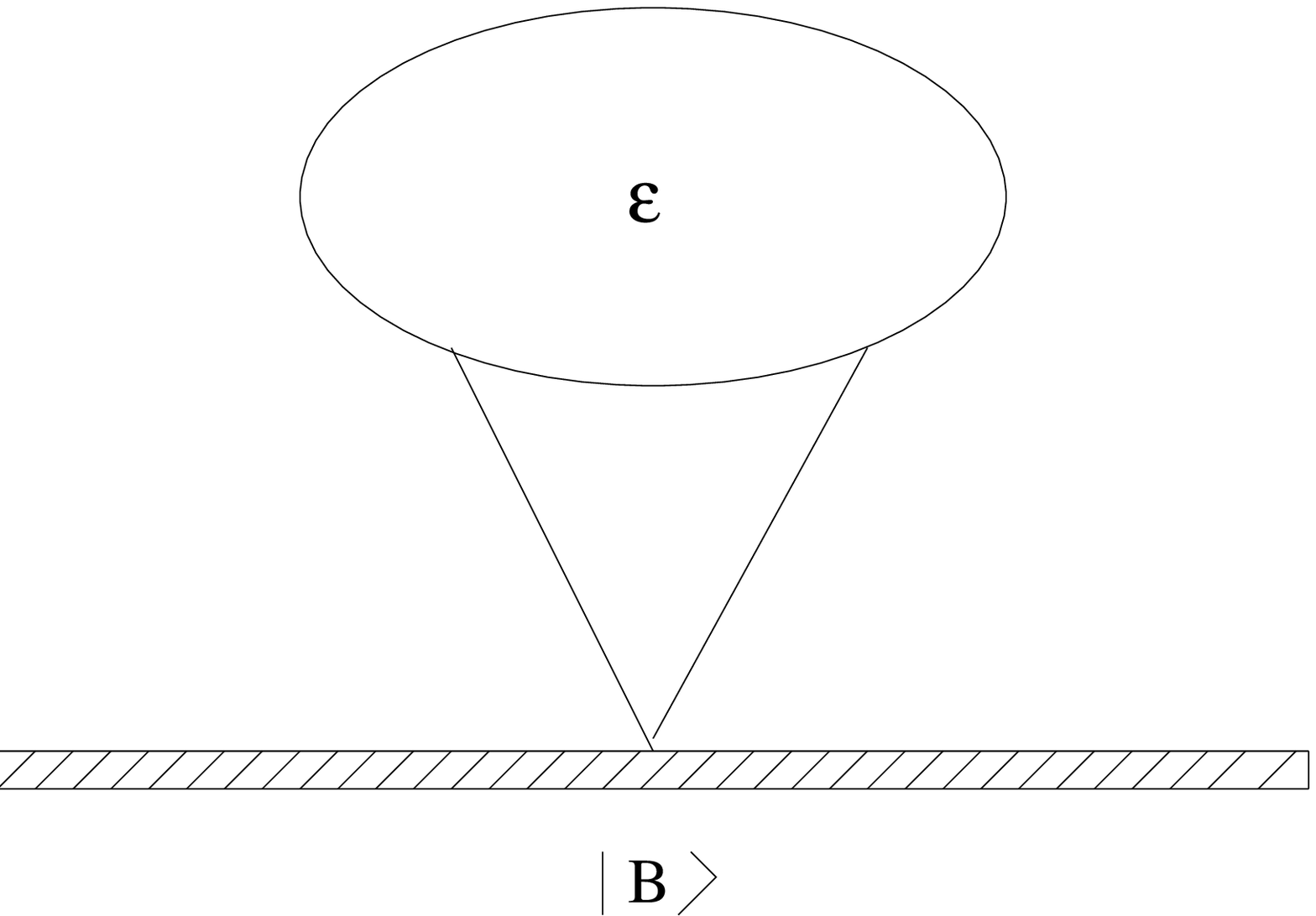}}}
\vskip .3in
{\leftskip .5in \rightskip .5in \noindent \ninerm \baselineskip=10pt
{}~~~~~~~~~~~~~~~Figure 2. One-point function of the energy operator.
\smallskip}} \bigskip
\endinsert
%%%%%%%%%%%%%%%%%%% End fig 2   %%%%%%%%%%%%%%%%%%%%%%%

The one-point function does not depend on $x$, as a consequence of
the translation invariance along this axis. The above
integral reduces to closed expressions in terms of the modified Bessel
functions in the case of free and fixed boundary conditions. Those are given
respectively by
\eqn\ebess{
\epsilon_0^{(\pm)}(t)\,=\,\mp m
\,\left[K_1(2mt) \pm K_0(2mt)\right]
.}
In the short distance limit $m t\rightarrow 0$, they have the
scaling form $ \epsilon_0^{(\pm)}(t) \sim \mp\frac{1}{2 t}$, in agreement
with the Conformal Field Theory prediction \ref\cl{J.L Cardy and D.C.
Lewellen, Phys. Lett. B 259 (1991), 274.}.
In the large distance limit, they instead decay exponentially with an extra
power term, which is different in the two cases, i.e.
\eqn\aymp{\eqalign{
\epsilon_0^{(+)}(t) & \sim - m  \,\sqrt{\frac{\pi}{mt}} \,e^{-2 mt} \,\, , \cr
\epsilon_0^{(-)}(t) & \sim \,\, \frac{m}{8}\,
\sqrt{\frac{\pi}{(mt)^3}} \,e^{-2 mt} \,\,\, . }}

Let us turn our attention to the one-point function of the energy operator
in the presence of a boundary magnetic field. This is given by
\eqn\eIxvi{
\epsilon_0(t,h) = -m  \int_0^{\infty} d\beta (\cosh\beta -1)
\frac{\cosh\beta + \kappa}{\cosh\beta - \kappa} e^{-2 m t \cosh\beta} \,\,\, .
}
By varying $h$, we may switch between the free and the fixed boundary
conditions, realizing therefore a Renormalization Group flow on the boundary.
At first sight, however, it seems impossible to interpolate between the
two functions $\epsilon_0^{(\pm)}(t)$, which present quite
a different behaviour.
(Notice, in particular, that the function $\epsilon_0^{(-)}(t)$ diverges as
$+\infty$ at $t\rightarrow 0$ whereas the other correlation function
$\epsilon_0^{(+)}(t)$ goes to $-\infty$
near the origin.)  To solve this apparent paradox, we have to correctly take
into account the interplay between the two variables $t$ and $h$. To this aim,
it is convenient to write down a differential equation satisfied by
$\epsilon_0(t,h)$. Observe that eq. \eIxvi\ can be initially expressed as
\eqn\eIxvii{
\epsilon_0(r,h) = -m
\int_0^{\infty} d\beta \int_0^{\infty} d\alpha \,
(\cosh\beta -1) (\cosh\beta + \kappa)
e^{- r \cosh\beta - \alpha (\cosh\beta - \kappa)}\,\, ,
}
($r\equiv 2 m t$) and the resulting integral can be expressed in terms of
modified Bessel function as
\eqn\eIxviii{
\epsilon_0(r,h) = -m e^{-\kappa r} \int_r^{\infty} d\eta\,
e^{\kappa\eta} \left[\frac{K_1(\eta)}{\eta} + (1-\kappa)(K_0(\eta) - K_1(\eta)
\right]\,\,\, .
}
Taking the derivative with respect to $r$ in both sides of this equation,
we obtain the following differential equation satisfied by $\epsilon_0(r,h)$
\eqn\eIxix{
\frac{\partial \epsilon_0(r,h)}{\partial r} + \kappa \,\epsilon_0(r,h)
= m\,\[
\frac{K_1(r)}{r} + (1-\kappa) (K_0(r) - K_1(r)) \] \,\,\, .
}
For $h=0$, using the differential equation satisfied by the Bessel functions,
we recover for $\epsilon_0(r,h=0)$ the previous result $\epsilon_0^{(+)}(t)$,
whereas $\epsilon_0^{(-)}(t)$ is obtained by taking directly the limit
$\kappa \rightarrow -\infty$ in the differential equation.

The differential equation is quite useful to analysing the short and
large distance behaviour of the one-point function $\epsilon_0(r,h)$.
For $r\rightarrow 0$, we can parameterize the solution as $\epsilon_0(r,h)
\sim C r ^{-\delta}$. Substituting into \eIxix, it is easy to see that
for {\it any } finite $\kappa$, the term which determines the short distance
behaviour in the RHS is $ m \frac{K_1(r)}{r}$, therefore
$
C = - m
$
and
$
\delta=1
$,
{\it independent} of the value of the magnetic field. Hence, as far as
$h$ is finite, all the curves $\epsilon_0(r,h)$ follow at a short distance
scale the behaviour associated to the free boundary condition.

For $r\rightarrow \infty$, we look for a solution of the form
$
\epsilon_0(r,h) \sim e^{-r} r^{-\omega} \sum_{k=0}^{\infty} a_k r^{-k}
$.
Substituting into \eIxix, using the large distance expansion of the
Bessel functions, and comparing the power series we have then the following
cases:

(i) for $\kappa=1$, $a_0 = -2m \sqrt{\frac{\pi}{2}}$
and $\omega = \frac{1}{2}$, so that
$\epsilon_0(r,h=0) \sim a_0 e^{-r}\,r^{-1/2}$.

(ii) for $-1 < \kappa < 1$, we have
$
a_0 = -m\sqrt{\frac{\pi}{2}} \frac{1+\kappa}{1-\kappa}
$
and $ \omega = \frac{3}{2}$, so that $\epsilon_0^{(h)}(r)
\sim a_0 e^{-r}\,r^{-3/2}$. In this range $a_0 < 0$ and the function
approaches the real axis from below.

(iii) for $\kappa=-1$ the first leading term of the expansion vanishes
and for the next leading term we have
$ a_1 = -m \frac{27}{16} \sqrt{\frac{\pi}{2}}$, $\omega = \frac{3}{2}$
and the corresponding solution goes to zero much faster,
$\epsilon_0(r,h=2\sqrt{m}) \sim a_1 e^{-r}\,r^{-5/2}$.

(iv) for $\kappa < -1$, we have
$
a_0 = -m \sqrt{\frac{\pi}{2}} \frac{1+\kappa}{1-\kappa}
$, $ \omega = \frac{3}{2} $
and $\epsilon_0^{(h)} \sim a_0 e^{-r}\,r^{-3/2}$. In this range
$a_0 > 0$ and the function approaches the real axis from above.

The one-point functions relative to some of the above cases are shown
in Figure 3.

%%%%%%%%%%%%%%%%%%%  fig 3  %%%%%%%%%%%%%%%%%%%%%
\midinsert
\epsfxsize = 3in
\bigskip\bigskip\bigskip\bigskip
\vbox{\vskip -.1in\hbox{\centerline{\epsffile{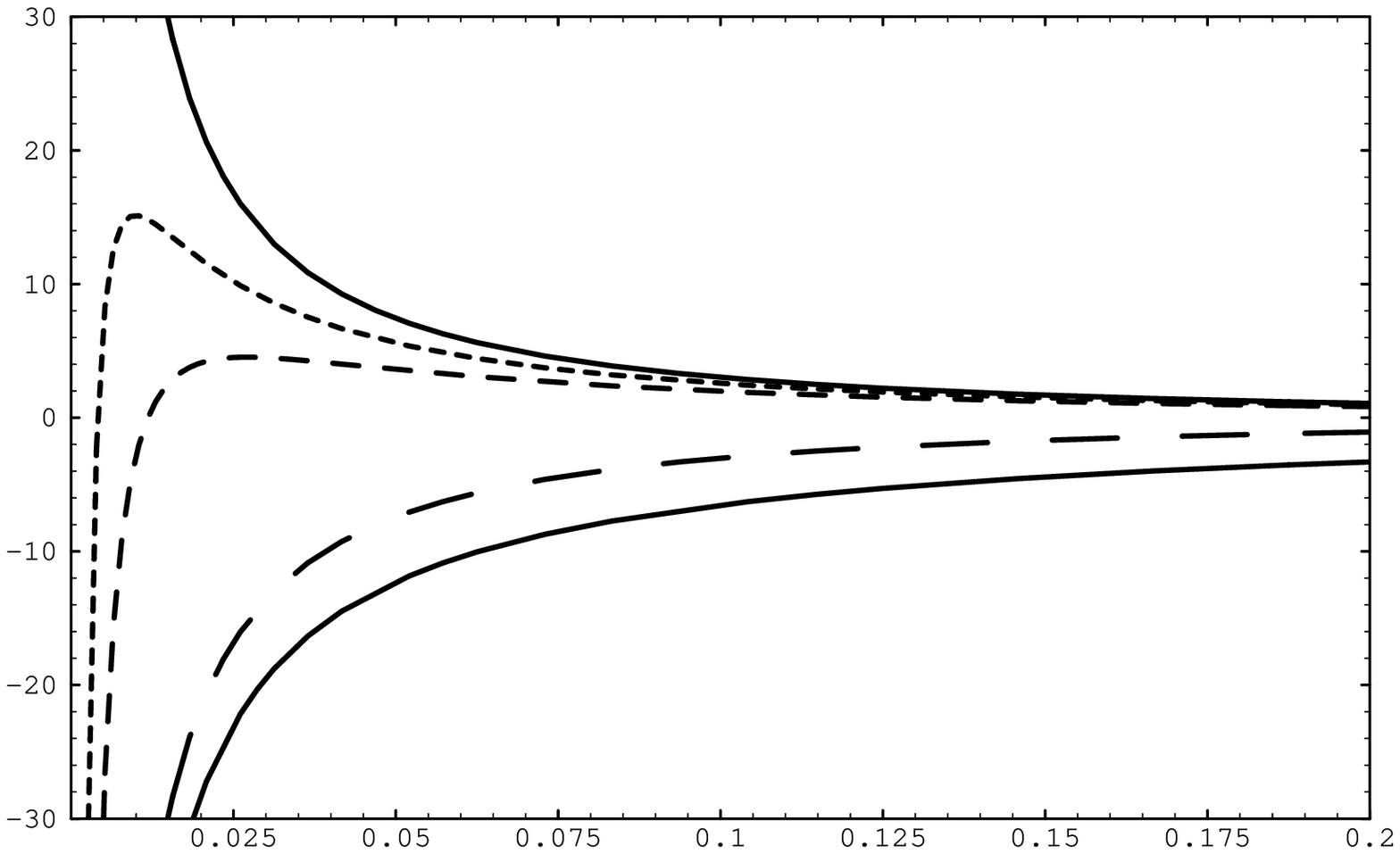}}}
\vskip .1in
{\leftskip .5in \rightskip .5in \noindent \ninerm \baselineskip=10pt
Figure 3. One-point function of the energy
operator versus $(mt)$ for different values of the boundary magnetic
field. Upper full line: fixed b.c.. Lower full line: free b.c..
Long dashed line: $h < h_c$. Short dashed lines: curves with $h > h_c$.
\smallskip}} \bigskip
\endinsert
%%%%%%%%%%%%%%%%%%% End fig 3   %%%%%%%%%%%%%%%%%%%%%%%

The result of this analysis is that for the boundary magnetic field smaller
than the critical value $h_c = 2 \sqrt{m}$, the one-point function does not
have any zero at finite value of $r$ whereas, for $h > h_c$, it crosses the
horizontal axes at a finite value of $r$, and its large $r$ behaviour
closely follows the behaviour relative to fixed boundary condition.

The above picture is consistent with a Renormalization Group analysis.
In fact, the boundary magnetic field is a relevant operator which therefore
cannot affect the behaviour of the correlation functions near the boundary,
as far as it assumes finite values. Hence, sufficiently close to $r=0$
the boundary always appears subject to free boundary condition. The
boundary magnetic field, however, influences the observables at large distance
scales and when $h > h_c$, the boundary always appears as subject to the fixed
boundary condition for an observer placed at $r \rightarrow \infty$. In an
intermediate scale, there is a non-trivial crossover between the two
different behaviours, which becomes rather sharp with increasing
$h$. The function $r(h)$, implicitly defined as the zeros
of the one-point function $\epsilon_0(r(h),h) = 0$, may be interpreted as a
phase diagram of the theory which divides the free and the fixed boundary
condition regions (Fig. 4).

%%%%%%%%%%%%%%%%%%%  fig 4  %%%%%%%%%%%%%%%%%%%%%
\midinsert
\epsfxsize = 3in
\bigskip\bigskip\bigskip\bigskip
\vbox{\vskip -.1in\hbox{\centerline{\epsffile{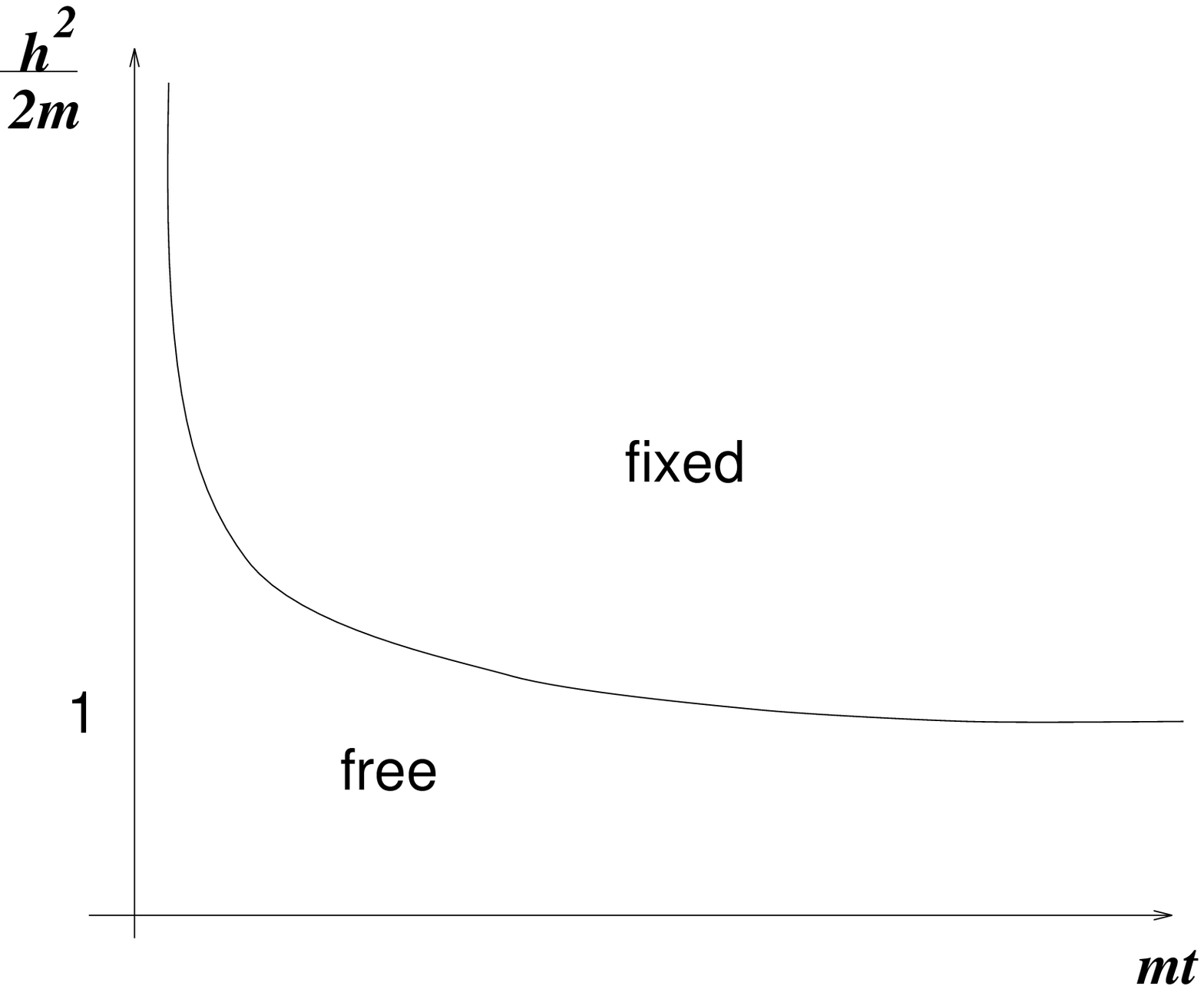}}}
\vskip .3in
{\leftskip .5in \rightskip .5in \noindent \ninerm \baselineskip=10pt
{}~~~~~~~~~~~~~~~~~Figure 4. Zeros of the one-point function
$\epsilon_0(r(h),h)$.
\smallskip}} \bigskip
\endinsert
%%%%%%%%%%%%%%%%%%% End fig 4   %%%%%%%%%%%%%%%%%%%%%%%

It is interesting to study the massless limit of the one-point function
\eIxvi. This corresponds to the physical situation of a critical
Ising model in the bulk but in the presence of a boundary condition which
breaks the conformal invariance of the model. With the change of variable
$y = 2 m t \cosh\beta$, the integral \eIxvi\  becomes
\eqn\eIxxviii{
\epsilon_0(t,h) = \frac{1}{2t} \int_{2 m t}^{\infty} dy e^{-t}
\sqrt{\frac{y - 2 m t}{y + 2 m t}}
\,\,\frac{y+ 2 m t \kappa}{y - 2 m t \kappa}
}
In the limit $m\rightarrow 0$, we have $ 2 m t \kappa \rightarrow -h^2 t
\equiv - z$
and the one-point function of the energy operator can be expressed as
\eqn\eEi{
\epsilon_0(t,h) = \frac{1}{2t} \left[1 + 2 z e^{z} Ei(-z)\right]
,} where $Ei(z)$ is the exponential integral function.

Let us consider now the connected two-point function of the energy operator in
presence of boundary magnetic field,
\eqn\twopoint{
G_c(x_1,t_1;x_2,t_2;h) = <\epsilon(x_1,t_1) \epsilon(x_2,t_2)>
- \epsilon_0(t_1,h)\, \epsilon_0(t_2,h) \,\,\, .}
By translation invariance along the $x$ axis, this function depends on the
difference $x=x_2 - x_1$. On the other hand, the presence of the boundary in
the time direction implies a dependence of $G$ on the two variables
$t_1$ and $t_2$. It is convenient to define $t\equiv t_2 - t_1$,
$\bar t \equiv= t_2 + t_1$ and $r\equiv \sqrt{x^2 + t^2}$. The correlation
function \twopoint\ can be easily computed by using eqs. \eIi\ and the form
factor \eIviii\ . Its closed form is given by
\eqn\finaltwo{\eqalign{
& G_c(x_1,t_1;x_2,t_2;h) = m^2 \[
\(\frac{\partial}{\partial (m x)} K_0(m r) +
F(m x,m\bar t)\)^2
+ \(\frac{\partial}{\partial (m t)} K_0(m r)\)^2 \right.\cr
& \left.~~~~~~~~~~~~~~~
- \(\frac{\partial}{\partial (m \bar t)} F(m x,m \bar t)\)^2
-\(K_0(m r) + \frac{\partial}{\partial (m x)} F(m x,m \bar t)\)^2 \]\,\, ,}}
where we have introduced the function
\eqn\auxiliar{
F(m \rho,m \tau) \equiv -\frac{i}{2} \int_{-\infty}^{+\infty}
dq  \frac{\cosh q + \kappa}{\cosh q - \kappa}
\, \tanh\frac{q}{2}\,\, \exp(- m \tau \cosh q + i m \rho \sinh q) \,\,\, .}
In the massless limit, the two-point function can be expressed as
\eqn\twomassless{
G_c(x_1,t_1;x_2,t_2;h) = \frac{1}{r^2} + {\cal F}(x,\bar t;h) \,
{\cal F}(-x,\bar t;h) \,\,}
where
\eqn\auxmassless{
{\cal F}(\rho,\tau;h) = \int_0^{\infty} dq \, \frac{2 q-h^2}{2q + h^2}
\, \exp(-\tau q + i \rho q) \,\,\, .}
It is easy to see that for free and fixed boundary conditions, eq.
\twomassless\ correctly reduces to the expression obtained by methods of
boundary conformal field theory \cardy\ .

\vskip .3in

\newsec{One-point function of the magnetization operator}

Let us consider now the one-point correlation functions of the Ising
magnetization field $\sigma (x,t)$. This can be computed by using the
formula
\eqn\sig{\sigma_0(t) = \sum_{n=0}^{\infty} \langle 0 | \sigma(x,t) | n\rangle
\langle n | B\rangle \,\,\, ,}
and by translation invariance, it only depends on $t$.
We denote by $\sigma_0^{(+)}(t)$
the one-point function relative to the free b.c., by $\sigma_0^{(-)}(t)$ the
one-point function with fixed b.c., and finally by $\sigma_0(t,h)$ the
one-point function in the presence of an arbitrary boundary magnetic field
$h$. In order to compute these one-point functions, we need the bulk Form
Factors of the magnetization operator \foot{With our normalization of the
Form Factors, the conformal limit of the two-point function in the bulk is
given by $\langle \sigma(r)
\sigma(0)\rangle = {\cal F}^{2}\,r^{-1/4}$, where
${\cal F} = 2^{-1/12} e^{1/8} A^{-3/2} \,m^{-1/8} = 0.73642778 \,
m^{-1/8}$, with $A=1.282427$ the Glasher constant. },
given by \kw \zcm :
\eqn\eIIiii{
\langle 0 | \sigma(0) |
\be_1 \cdots \be_{2n} \rangle  =  \,i^n \prod_{i<j}
\tanh \( \frac{\be_i - \be_j }{2}  \) \,\,\,. }
Since the boundary state $|B \rangle$ consists of pairs of particles
with opposite rapidity, we have to specialize the previous formula and
evaluate
\eqn\eIIiv{\eqalign{
\langle 0 | \sigma (0) | - \be_1 , \be_1 , \cdots - \be_n , \be_n \rangle
& = (-i)^n\, \,   \prod_{i=1}^n  \tanh \be_i \,
\prod_{i<j}
\( \tanh \frac{\be_i - \be_j }{2}  \tanh \frac{\be_i + \be_j}{2}  \)^2 =
\cr
& = (- i)^n \, \prod_{i=1}^n  \tanh \be_i \,
\prod_{i<j}
\(  \frac{ \cosh \be_i  - \cosh \be_j }{\cosh \be_i + \cosh \be_j } \)^2
\,\,\, .} }
Noting that
\eqn\eIIvi{
\prod_{i<j}
\(  \frac{ \cosh \be_i  - \cosh \be_j }{\cosh \be_i + \cosh \be_j } \)^2 \,
= \,{\rm det} \( \frac{2 \sqrt{ \cosh \be_i \cosh \be_j } }{\cosh \be_i
+ \cosh \be_j }\) \equiv {\rm det}\, \, W(\beta_i,\beta_j) \,\,
, }
we can express the one-point function $\sigma_0^{(\pm)}(t)$ as a Fredholm
determinant, namely,
\eqn\eIIvii{\eqalign{
\sigma_0^{(\pm)} (t) & =
\, \sum_{i=0}^\infty \inv{n!}
\int_0^{+\infty}\frac{d\beta_1}{2\pi}\cdots \int_0^{+\infty}
\frac{d\beta_n}{2\pi}
\(\prod_{k=0}^{n} \(-i \tanh\beta_k \,\hat R_{\pm}(\beta_k)\)\) \,{\rm det}
W(\beta_i,\beta_j) = \cr
 & = \, {\rm Det} ( 1 -z_{\pm}\, V_{(\pm)}(t)) \,\, ,}}
where the kernels are given by
\eqn\eIIviii{
V_{(\pm)}(\beta_i,\beta_j,t)  =\,\frac{e^{(\pm)}(\beta_i,t) \,
e^{(\pm)}(\beta_j,t) }
{\cosh\beta_i + \cosh\beta_j}\,\,  , ~~~~~
e^{(\pm)}(\beta,t) = \sqrt{\cosh\beta \pm 1} \, e^{-mt \cosh\beta}
{}.
}
In the above formula, the physical correlator is obtained for
$z_{\pm} = \pm 1/(2\pi)$,  but it is convenient to regard the
above expressions as functions of the parameters $z_{\pm}$.

The one-point function $\sigma_0(t,h)$ can also be cast into a
Fredholm determinant form as
\eqn\eIIviiv{
\sigma_0(t,h) = \, {\rm Det} ( 1 -z\, V(t,h)) \,\, ,  }
where the kernel is now given by
\eqn\eIIviiiv{\eqalign{
V(\beta_i,\beta_j,t,h)  &=\,\frac{E(\beta_i,t,h) \,
E(\beta_j,t,h) }
{\cosh\beta_i + \cosh\beta_j} \,\, ,   \cr
E(\beta,t,h) &= \sqrt{(\cosh\beta -1)
\frac{\cosh\beta +\kappa}{\cosh\beta -\kappa}} \,\, e^{-mt \cosh\beta}
\,\,\, .
\cr}}
In this case, the physical correlator is obtained for $z = 1/(2 \pi)$.

It is quite easy to show that the crossover phenomenon previously
discussed for the one-point function of the energy operator also takes place
for $\sigma_0(t,h)$. Let us initially consider the short
distance behaviour of  eqs. \eIIvii\   and \eIIviiv. As proved in
appendix A, in the limit $m t \rightarrow 0$, we have
\eqn\sd{\eqalign{
\sigma_0^{+}(t) & \sim (2 t)^{3/8} \,\, , \cr
\sigma_0^{-}(t) & \sim (2t) ^{-1/8} \,\,\, .
}}
Concerning the behaviour of the one-point function $\sigma_0(t,h)$ in the limit
$m t \rightarrow 0$, its power law singularity is {\it independent} of $h$,
as far as $h$ is finite, and coincides with that one relative to the free
boundary condition, i.e. $\sigma_0(t,h) \sim (2t)^{3/8}$. Hence, varying $h$,
all the curves $\sigma_0(t,h)$ follow at short distance the power-law behaviour
dictated by the free boundary conditions.

Let us evaluate now the large distance behaviour. For the fixed boundary
conditions, we have
\eqn\larfix{
\sigma_0^-(t) = \,\[ 1 + \frac{1}{16 \sqrt{\pi}} \frac{e^{-2 m t}}
{(m t)^{3/2}} + \cdots\] \,\, , }
i.e. this function approaches the bulk expectation value from above. For the
free boundary condition, we have instead
\eqn\larfree{
\sigma_0^+(t) = \, \[1 -
\frac{1}{2 \sqrt{\pi}} \frac{e^{-2 m t}}{(mt)^{1/2}}
+ \cdots \] \,\, , }
i.e. the bulk expectation value is reached from below. In the presence of
a boundary magnetic field, the asymptotical approach to the bulk vacuum
expectation value is given for $\kappa \neq -1$ by the expression
\eqn\larmag{
\sigma_0(t,h) = \, \[1 - \frac{1}{16 \sqrt{\pi}}\,\frac{1+\kappa}{1-\kappa}
\frac{e^{-2 m t}}{(mt)^{3/2}} + \cdots \] \,\, , }
whereas for $\kappa = -1$
\eqn\larmagg{
\sigma_0(t,h) = \[1 - \frac{3}{128 \sqrt{\pi}} \frac{e^{-2 mt}}
{(m t)^{5/2}} + \cdots \] \,\,\, . }
{}From \larmag\ we see that for $-1 < \kappa < 1$, the curve reaches its
asymptotic behaviour from below, whereas for $\kappa < -1$ from above. At
$\kappa =-1$ there is a faster decreasing behaviour. Hence, also
for the one-point function of the magnetization operator there is the
typical Renormalization Group crossover between free and fixed boundary
conditions, moving away from the boundary (Fig. 5).

%%%%%%%%%%%%%%%%%%%  fig 5  %%%%%%%%%%%%%%%%%%%%%
\midinsert
\epsfxsize = 3in
\bigskip\bigskip\bigskip\bigskip
\vbox{\vskip -.1in\hbox{\centerline{\epsffile{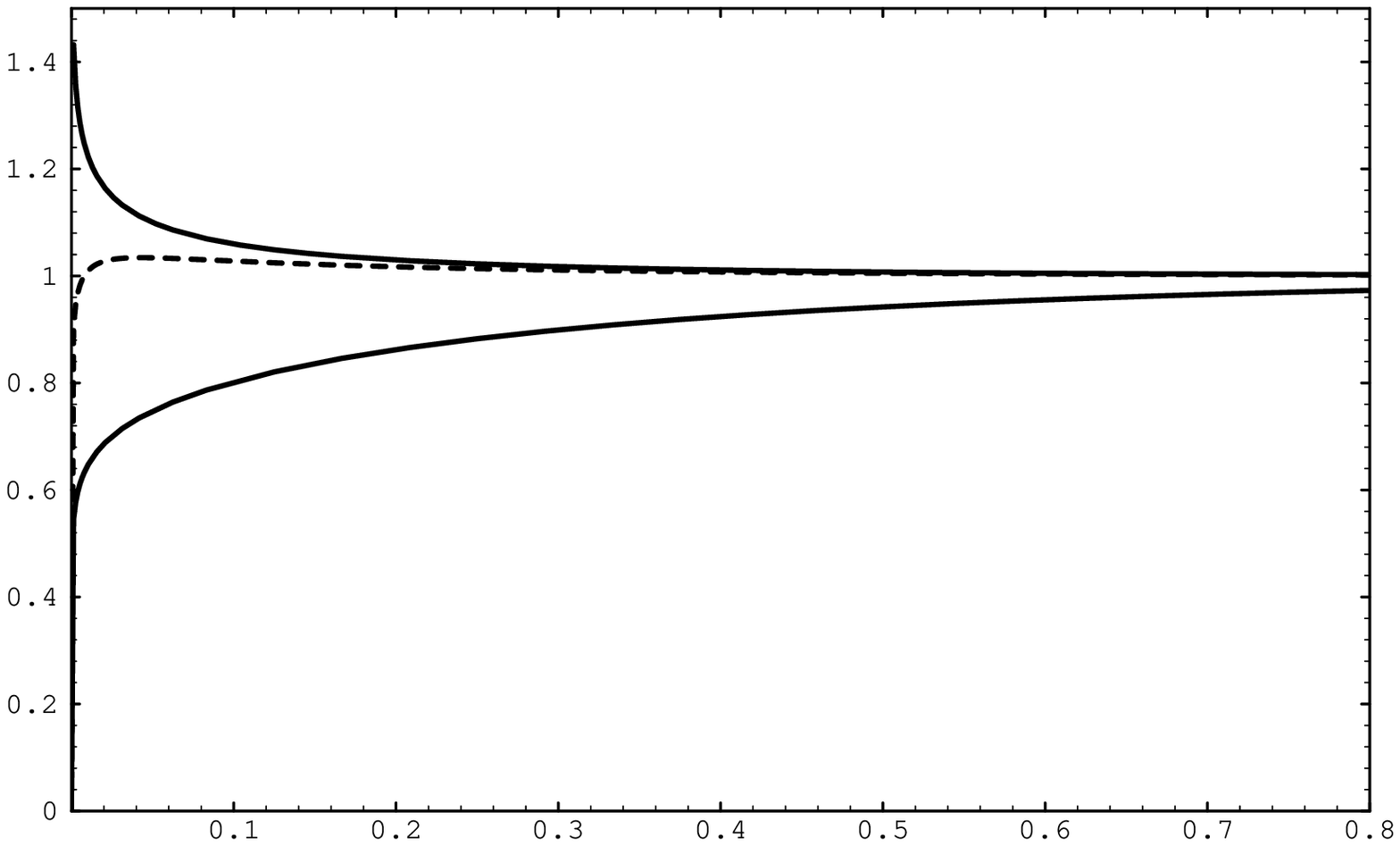}}}
\vskip .1in
{\leftskip .5in \rightskip .5in \noindent \ninerm \baselineskip=10pt
Figure 5. One-point function of the magnetization operator versus $(mt)$.
Upper full line: fixed b.c.. Lower full line: free b.c.. Dashed line: curve
with $h> h_c$.
\smallskip}} \bigskip
\endinsert
%%%%%%%%%%%%%%%%%%% End fig 5   %%%%%%%%%%%%%%%%%%%%%%%

\newsec{Differential equation for the magnetization one-point function}

Bariev has shown that for the free and fixed boundary condition,
the one-point functions of the magnetization can be expressed
in terms of a solution to the Painlev\'e differential equation \bariev~.
He started from a lattice description,
from which he obtained the same Fredholm determinant expressions
as we directly obtained by using the form factor approach. In this section we
derive differential equations for arbitrary magnetic field $h$
using the general techniques developed in \ref\rkor{A. R. Its, A. G. Izergin,
V. E. Korepin, and N. A. Slavnov, Int. J. Mod. Phys. B4 (1990) 1003.}
\ref\rkorb{V. E. Korepin, N. M. Bogoliubov and A. G. Izergin,
{\it Quantum Inverse Scattering Method and Correlation
Functions}, (Cambridge Univ. Press, Cambridge, 1993).}.
As we will show, these will turn out to be coupled non-linear
partial differential equations in the variables $t$ and $\kappa$.

Making the change of variable $u \equiv \cosh\beta$, we can initially
write the one-point function $\sigma_0(t,h)$ as
\eqn\fff{
\langle 0 |\sigma(t) | B(h)\rangle =
{\rm det} (1 + {\cal V}) =
\sum_{n=0}^{\infty} \frac{1}{n!}\,\int_1^{+\infty} \frac{du_1}{2\pi} \cdots
\int_1^{+\infty} \frac{du_n}{2\pi} \,\, {\rm det} {\cal V}(u_i,u_j,t) }
where
\eqn\kernel{\eqalign{
{\cal V}(u_i,u_j,t)  &=\,\frac{e(u_i,t) \,
e(u_j,t) }
{u_i + u_j}   \,\, ,\cr
e(u,t) &= \(\frac{u-1}{u+1}\)^{1/4} \, \sqrt {\frac{\kappa + u}{\kappa -u}}
\, e^{-mt u}
\,\,\, .
}}
One first needs the resolvent $R_\pm$, defined as follows:
\eqn\eIIx{
(1-R_\pm) (1 \pm {\cal V} ) = 1\,\,\, .}
In the following, it will always be implied that Fredholm operators
are multiplied in the usual way, e.g.
\eqn\eIIxb{
(R V)(u,v) = \int_1^{+\infty} \frac{dw}{2\pi}  R(u,w) V(w, v ) \,\, , }
and similarly for multiplication  of operators with functions:
\eqn\eIIxc{
(R e)(u) = \int_1^{+\infty} \frac{dv}{2\pi}  R(u,v) e(v) \,\,\, . }
Resolvents for the kernels of the type \eIIviii\ were described
in \ref\rbl{D. Bernard and A. LeClair, Nucl. Phys. B426 [FS]
(1994) 534.}.  Define functions $f^\pm (u)$ by the formulas
\eqn\eIIxi{
(1 \pm {\cal V} ) f^\pm = e   ~~~\Longleftrightarrow ~~~
f^\pm = (1-R_\pm ) e \,\,\, .}
Then
\eqn\eIIxii{
R_\pm (u,v) = H(u,v) \pm F(u,v) \,\, , }
where
\eqn\eIIxiii{\eqalign{
H(u,v) &= \frac{f^- (u) f^+ (v) - f^+ (u) f^- (v) }{2(u-v)} \,\, ,\cr
F(u,v) &= \frac{f^+ (u) f^- (v) + f^- (u) f^+ (v) }{2(u+v)} \,\,\, .\cr
}}

The next step is to identify variables with the property that
derivatives of the kernel ${\cal V}$ with respect to such variables
acts as a projector, i.e. removes the $1/(u+v)$ factor.
The spacial variable $t$ obviously has this property:
\eqn\eIIxiv{
\d_t {\cal V}(u,v) = - m\, e(u) e(v) \,\,\, .}
One also has
\eqn\eIIxv{
\d_\kappa {\cal V}(u,v) = (uv - \kap^2 ) e'(u) e'(v) \,\, , }
where
\eqn\eIIxvi{
e' (u) = \frac{e(u)}{u^2 - \kap^2 } \,\,\, .}
For future convenience, define also
\eqn\eIIxvib{
\hat{e} (u) = \frac{u}{u^2 - \kap^2}  e(u) \,\,\, .}

The rest of the derivation of correlation functions parallels
closely the treatment of Ising 2-point functions given in
\rbl\ with $z, \zb$ replaced by $r, \kap$.
Let
\eqn\eIIxvii{
\tau_\pm = {\rm Det} (1 \pm {\cal V}) \,\,\, . }
Then
\eqn\eIIxviii{\eqalign{
\d_t \log \tau_\pm
&= \d_t {\rm Tr} \log (1 \pm {\cal V} ) =  {\rm Tr} (1 - R_\pm ) \d_t
{\cal V} \cr
&= \mp m\, P_\pm \,\, , \cr}}
where
\eqn\eIIxix{
P^\pm = <e, f^\pm > \equiv \int \frac{du}{2\pi} e(u) f^\pm (u) \,\,\, .}

Next, we compute $\d_\kap P_\pm$, for which we need
$\d_\kap f^\pm$.  Taking the derivative of \eIIxi, one obtains
\eqn\eIIxx{
(1\pm {\cal V}) \d_\kap f^\pm \pm \d_\kap {\cal V} f^\pm = \eh \,\,\, .}
The latter implies
\eqn\eIIxxi{
(1\pm {\cal V}) \d_\kap f^\pm \mp \kap^2 e' <e' , f^\pm >
\pm \eh < \eh , f^\pm >  = \eh \,\,\, .}
Define the additional functions
\eqn\eIIxxii{
\fh^\pm = (1-R_\pm) \eh\,\, , ~~~~~~~~f'^\pm = (1-R_\pm ) e' \,\,\, .}
Then multiplying \eIIxxi\ by $(1-R_\pm )$, one obtains
\eqn\eIIxxiii{
\d_\kap f^\pm = \fh^\pm \pm \kap^2 D^\pm f'^\pm \mp
\dh^\pm \fh^\pm \,\, , }
where
\eqn\eIIxxiv{
\eqalign{
D^\pm &= <e' , f^\pm > = <e, f'^\pm > \,\, ,\cr
\dh^\pm &= <\eh , f^\pm > = <e, \fh^\pm > \,\,\, .\cr
}}
One then finds
\eqn\eIIxxv{\eqalign{
\d_\kap P^\pm
&= <\d_\kap e , f^\pm > + <e, \d_\kap f^\pm > \cr
&= 2 \dh^\pm \pm \kap^2 (D^\pm )^2 \mp (\dh^\pm )^2 \,\,\, .\cr}}

We can also
obtain expressions for $\d_t D^\pm $ and $\d_t \dh^\pm $.  For
this, one needs $\d_t f^\pm$.  This is the same as in \rbl:
\eqn\eIIxxvi{
\d_t f^\pm = m\,\[\pm ( P^+ + P^- ) f^\pm  - u f^\mp \]\,\,\, . }
One then has
\eqn\eIIxxvii{\eqalign{
\d_t D^\pm  &= < \d_t e' , f^\pm > + <e' , \d_t f^\pm > \cr
&= m\,\[ \pm ( P^+ + P^- ) D^\pm  - \dh^+ - \dh^- \]\, . \cr}}
Similarly,
\eqn\eIIxxviii{
\d_t \dh^\pm = m\,\[\pm ( P^+ + P^- ) \dh^\pm  - (P^+ + P^- )
- \kap^2 (D^+ + D^- )\]\, . }

Finally, there is one non-trivial constraint among the
$D, \dh$ potentials, which is proven in the appendix B:
\eqn\eIIxxix{
\dh^+  = \dh^- - \dh^+ \dh^- + \kap^2 D^+ D^-  \,\,\, . }

We can summarize the above results as follows.  Define
\eqn\eIIxxx{\eqalign{
P  &= P^+ + P^- \,\, , ~~~~~~~~~~~~~~~Q = P^+ - P^-  \,\, ,\cr
C &= 2 + \dh^- - \dh^+ \,\, , ~~~~~~~~~~\tilde{C} = -( \dh^+ + \dh^- )
\,\, ,\cr
D &= D^+ + D^- \,\, ,~~~~~~~~~~~~~~~\dt = D^+ - D^- \,\,\, .\cr
}}
Then,
\eqn\eIIxxxi{\eqalign{
& \d_t  \log (\tau_+ / \tau_- )  =  - m P \,\, ,\cr
& \d_t  \log (\tau_+  \tau_- )   = - m Q \,\, , \cr
& \d_\kap P = \kap^2 D \dt - C \ct \,\, ,\cr
& \d_\kap Q = 2 - \inv{2} \( C^2 + \ct^2 \) + \frac{\kap^2}{2} \( D^2 + \dt^2
\)
\,\, ,\cr}}
and
\eqn\eIIxxxii{
\eqalign{
\d_t D & = m\,\[2 \ct + P \dt \] \,\, ,\cr
\d_t \dt & = m\, P D \,\, , \cr
\d_t C & = m\, P \ct  \,\, ,\cr
\d_t \ct & = m\, \[P C + 2 \kap^2 D\]\,\,\, . \cr }}
The constraint reads
\eqn\eIIxxxiii{
C^2 - \ct^2 + \kap^2 (D^2 - \dt^2 ) = 4 \,\,\, . }

The equations \eIIxxxi\ indicate that $\d_t \d_\kap \log (\tau_\pm )$
can be expressed in terms of the four functions $C, \ct, D , \dt$.
In turn, the function $P$ can be eliminated from \eIIxxxii\ to
obtain differential equations involving  $C, \ct, D , \dt$ only.
For instance, expressing $P = \d_t \dt /mD$, and using the expression
for $\d_\kap P$, one finds:
\eqn\einst{
D \d_\kap \d_t \dt - \d_t \dt \d_\kap D = m D^2 ( \kap^2 D \dt - C \ct ) .}
Three additional differential equations of the same kind can
be similarly obtained by eliminating $P$ from each of the
equations in \eIIxxxii.

\newsec{Concluding Remarks}

In this paper we have studied the effects induced by a boundary
Renormalization Group flow on the correlation functions of the
two-dimensional Ising model. Using the form factors of the local
energy and magnetization operators, we have computed their one-point
functions in the presence of a boundary magnetic field as sums on the
intermediate states of the bulk theory. The integral representations of
the correlation functions can be used to derive the differential
equations which they satisfy. In view of the ``free'' nature of the
energy operator in the Ising model, its one-point functions in the presence
of a boundary magnetic field is a solution of a linear non-homogeneous
differential equation, given in \eIxix . On the contrary, the magnetization
operator is an interacting field and for its one-point function in the
presence of a boundary magnetic field, we obtain the system of non-linear
differential equations \eIIxxxi , \eIIxxxii ,with the constraint expressed
by \eIIxxxiii . The rather different mathematical structure of these
differential equations reflects the distinct dynamics of those fields.

Some simplications in the equations are expected in the massless limit of
the theory. This corresponds to the Ising model with an infinite correlation
length in the bulk but with a violation of the conformal symmetry induced by
the boundary interaction. We have analysed this interesting situation
for the cases of the one-point and the two-point functions
of the energy operator. Concerning the one-point function of the
magnetization operator, it is known that it satisfies
in the massless limit a linear differential equation of
hypergeometric type \ref\rchatz{R. Chatterjee
and A. Zamolodchikov,
{\it Local Magnetization in Critical Ising Model with Boundary Magnetic
Field}, hep-th/9311165.}.
It would be interesting to obtain directly the result of ref. \rchatz\
by analysing the massless limit of the system of non-linear differential
equations derived in this paper.

\bigskip\bigskip

\centerline{Acknowledgements}

\bigskip

We would like to thank V. Korepin for useful discussions.
This work is supported by an Alfred P. Sloan Foundation fellowship,
and the National Science Foundation in part through the National
Young Investigator program.

\bigskip\bigskip\bigskip

\appendix{A}{}
In this appendix we calculate the power law singularity of the one-point
function of the magnetization operator. Let us start by analysing the free and
fixed boundary conditions.
An equivalent expression for $\sigma_0^{(\pm)}(t)$ is given by
\eqn\cumu{\eqalign{
\sigma_0^{(\pm)}(t) & = {\rm det} (1 - z_{\pm} V_{(\pm)}) =
 \exp \[{\rm Tr} \log (1 - z_{\pm} V_{(\pm)})\] = \cr
& = \exp \[-\sum_{n=1}^{\infty} \frac{z^n_{\pm}}{n}
\int_0^{+\infty} d\beta_1 \cdots \int_0^{+\infty}  d\beta_n
\prod_{i=1}^n e^{-2 m t \cosh\beta_i} \frac{\cosh\beta_i \pm 1}
{\cosh\beta_i + \cosh\beta_{i+1}}\] = \cr
& \equiv \exp \[ - \sum_{n=1}^{\infty} \frac{z_{\pm}^n}{n} Z_{\pm}^{(n)}(t)
\]\,\, ,}}
(where $ \beta_{n+1} \equiv \beta_1$). Using now
\eqn\denom{
\frac{1}{\cosh\beta_i + \cosh\beta_{i+1}} =
\int_0^{+\infty} d\alpha \,\,e^{-\alpha (\cosh\beta_i + \cosh\beta_{i+1})}
\,\, ,}
the coefficients $Z_{\pm}^{(n)}(t)$ in \cumu\  can be expressed as
\eqn\coeff{\eqalign{
Z^{(n)}_{\pm}(t) & = \int_{mt}^{+\infty} d\alpha_1 \cdots
\int_{mt}^{+\infty} d\alpha_n \[K_1(\alpha_1 + \alpha_2) \pm
K_0(\alpha_1 + \alpha_2)\] \cr
& \times \[K_1(\alpha_2 + \alpha_3) \pm
K_0(\alpha_2 + \alpha_3)\] \cdots
\[K_1(\alpha_n + \alpha_1) \pm
K_0(\alpha_n + \alpha_1)\]  \,\, ,}}
where $K_1(x)$ and $K_0(x)$ are modified Bessel functions. For
$m t \rightarrow 0$, the leading singularity of these expressions comes from
\eqn\sing{
Z^{(n)}_{\pm}(t)  \sim \int_{mt}^{+\infty} d\alpha_1 \cdots
\int_{mt}^{+\infty} d\alpha_n K_1(\alpha_1 + \alpha_2)
K_1(\alpha_2 + \alpha_3) \cdots K_1(\alpha_n + \alpha_1)\,\,\, .}
It is easy to estimate the sum of the most singular terms. To do this,
let us express $\sigma_0^{(\pm)}(t)$ in terms of the eigenvalues
$\lambda_{\pm}(t)$ of the integral operators $V_{\pm}$ and their multiplicity
$a_i(t)$ as
\eqn\product{
\sigma_0^{(\pm)}(t)\,=\, \prod_{i=1}^{\infty}
\(1 - z_{\pm} \,\lambda_{\pm}^{(i)}(t)\)^{a_i(t)}
\,\,\, .
}
As far as $(mt)$ is finite, the kernel is square integrable.
However, when $(mt) \rightarrow 0$, the operator becomes
unbounded. In this limit, the eigenvalues becomes dense in the interval
$(0,\infty)$ according to the distribution
\eqn\dis{
\lambda(p)\,=\,\frac{2\pi}{\cosh\pi p} \,\, ,
}
whereas, from Mercer's theorem, their multiplicity grows logarithmically as
$a_i \sim \frac{1}{\pi} \ln\frac{1}{mx}$. Therefore
\eqn\shdis{
\log\sigma_0^{(\pm)}(t) = x_{\pm} \log t }
where the critical exponents $x_{\pm}$ of the
magnetization operator relative to free and fixed boundary conditions are
given by
\eqn\anom{
x_{\pm}\,= \,\frac{1}{\pi^2}\,
\int_0^{\infty} dp \,\ln\left(1 - \frac{2\pi z_{\pm}}{\cosh p}\right)
\,=\, \frac{1}{8} - \frac{1}{2\pi^2} \arccos^2(-2\pi z_{\pm})\,\,\,.
}
Substituting the values of $z_{\pm}$ we obtain $x=3/8$ for the free b.c. and
$x=-1/8$ for the fixed b.c.

Repeating the same analysis for the one-point function in the presence of the
boundary magnetic field $h$ by using the formula \cumu , it is easy to see
that, as far as $h$ is finite, the most leading singularity of the
corresponding $Z^{(n)}$ is always given by \sing\ and can be estimated
according to the equation \anom\  with $z=1/(2\pi)$, i.e. $x=3/8$.

\appendix{B}{}

Here, we relate the functions $\fh^\pm , f'^\pm $ to the
functions $f^\pm$ and prove the constraint \eIIxxix.

Using the identity
\eqn\eAo{
\inv{u+v} \inv{v+k}  = \inv{u-k} \( \inv{v+k} - \inv{u+v} \) \,\, , }
one first easily shows that
\eqn\eAi{\eqalign{
H(u,v) \inv{v+k}  &=
\inv{u+k} H(u,v)
+ \inv{2} \inv{u+k} \inv{v+k}
\( f^- (u) f^+(v) - f^+(u) f^- (v) \)\,\, ,  \cr
F(u,v) \inv{v+k}  &= -
\inv{u-k} F(u,v)
+ \inv{2} \inv{u-k} \inv{v+k}
\( f^+ (u) f^-(v) + f^-(u) f^+ (v) \) \,\,\, . \cr
}}
{}From the definition,
\eqn\eAii{
f'^\pm (u) = \( (1- R_\pm ) \frac{e(v)}{v^2 - \kap^2} \) (u) \,\,\, .}
Factorizing $v^2 - \kap^2 = (v+\kap) (v-\kap )$ in the above
equation, and applying \eAi\ twice, one obtains
\eqn\eAiii{
f'^\pm (u) = \inv{u^2 - \kap^2}
\[ (1\pm Q^\mp ) f^\pm  \mp \kap D^\mp f^\pm \mp u f^\mp D^\pm \]\,\, , }
where
\eqn\eAiv{
Q^\pm = <f^\pm , \frac{e(u)}{u-\kap} > = \dh^\pm + \kap D^\pm \,\,\, . }
Using the identities
\eqn\eAv{\eqalign{
H(u,v) v &= u H(u,v) + \inv{2}
\( f^+ (u) f^- (v) - f^- (u) f^+ (v) \) \,\, ,\cr
F(u,v) v &= -u F(u,v) + \inv{2}
\( f^+ (u) f^- (v) + f^- (u) f^+ (v) \)\,\, , \cr
}}
one also finds
\eqn\eAvi{
\fh^\pm (u) = u f'^\mp \mp D^\mp f^\pm \,\,\, . }

The constraint \eIIxxix\ follows from using \eAvi, \eAiii\ to express
$\fh^\pm$ in terms of $f^\pm$ in the equation
$\dh^\pm = <\eh , f^\pm> = <e, \fh^\pm > $.

\listrefs
\end